\documentclass[11pt]{article}

\usepackage{ amsmath, amssymb, bm }

\newcommand{\given}{\, | \,}

\begin{document}

\large

\textsf{Contribution to the discussion of ``Martingale posterior distributions," by Edwin Fong, Chris Holmes, and Stephen G.~Walker}

\vspace*{0.1in}

\normalsize

\textbf{David Draper} and \textbf{Erdong Guo} (\textit{University of California, Santa Cruz, USA})

We have two comments motivated by this interesting paper.

\begin{itemize}

\item[(1)]

The idea, introduced early in the paper, that ``the object of interest is fully defined
once all the observations have been viewed'' is almost exactly 100 years old: it was a cornerstone of the remarkable paper by Fisher (1922), and has been referred to for many decades as \textit{Fisher consistency}. We are surprised that the authors did not make this connection.

\item[(2)]
  
The authors make strong distinctions between the frequentist and Bayesian bootstraps. We would like to point out the not-so-widely-known fact that \textit{the frequentist bootstrap is actually an instance of Bayesian nonparametric inference}, as follows. Suppose that the context $\mathbb{ C }$ of the problem under study by You (Good, 1950: a person wishing to reason sensibly in the presence of uncertainty) implies that Your uncertainty about real-valued observables $\{ Y_1, Y_2, \dots \}$, which have not yet been observed, is exchangeable. Then de Finetti's Representation Theorem for real-valued outcomes tells us that this is equivalent to the Bayesian hierarchical model
\begin{eqnarray} \label{e:de-finetti-1}
( F \given \bm{ \mathcal{ B }^* } ) & \sim & p ( F \given \bm{ \mathcal{ B }^* } ) \nonumber \\
\left\{ \begin{array}{c} ( Y_i \given F \, \bm{ \mathcal{ B }^* } ) \\ ( i = 1, \dots, n ) \end{array} \right\} & \stackrel{ \textrm{IID} }{ \sim } & F \, ,
\end{eqnarray} 
in which $F$ is the empirical CDF based on $\{ Y_1, Y_2, \dots \}$, $n$ is a finite positive integer, and $\bm{ \mathcal{ B }^* }$ is a finite set of propositions, all rendered true by context $\mathbb{ C }$ and exhaustive of all relevant contextual information. As is well known, (a) the conjugate prior for $F$ in this model is the family $DP ( \alpha, F_0 )$ of Dirichlet processes, where $\alpha > 0$ and $F_0$ represent the appropriate prior sample size and prior estimate of $F$, respectively, based on Your information external to the observed data set $\bm{ y } = ( y_1, \dots, y_n )$, and (b) conjugate updating yields the posterior
\begin{equation} \label{e:de-finetti-2}
DP \left( \alpha + n, \frac{ \alpha \, F_0 + n \, \hat{ F }_n }{ \alpha + n } \right)
\end{equation}
for $F$, in which $\hat{ F }_n$ is the empirical CDF based on $\bm{ y }$. To create a low-information prior it is tempting to send $\alpha \downarrow 0$; Terenin and Draper (2017) have shown that this is mathematically meaningful, with the resulting prior, which they call $DP ( 0 )$, yielding the important-for-statistical-science posterior $DP ( n, \hat{ F }_n )$. A corollary of a result in Terenin, Magnusson, Jonsson, and Draper (2018) then yields the following theorem, stated informally:

\begin{quote}

\textbf{Theorem} (Draper and Guo, 2023) Under the conditions detailed above, frequentist bootstrap samples of size $n$ from $( y_1, \dots, y_n )$ are asymptotically stochastically indistinguishable from stick-breaking samples of the same size from $DP ( n, \hat{ F }_n )$.

\end{quote}

We find empirically that the frequentist bootstrap approximation is good to excellent even for $n$ as small as 25; this has useful implications for high-quality Bayesian data science.

\end{itemize}

\begin{center}

\textbf{References}

\end{center}

\small

\begin{itemize}

\item[ ]

Draper D and Guo E (2023). Optimal Bayesian Analysis in Digital Experimentation at Big-Data Scale: the Frequentist Bootstrap is Actually an Instance of Bayesian Nonparametric Inference. In preparation.

\item[ ]

Fisher RA (1922). On the Mathematical Foundations of Theoretical Statistics. \textit{Philosophical Transactions of the Royal Society of London, Series A}, \textbf{222} (594--604): 309--368.

\item[ ]

Good IJ (1950). \textit{Probability and the Weighing of Evidence.} London: Griffin.

\item[ ]

Terenin A and Draper D.~(2017). A Noninformative Prior on a Space of
Distribution Functions. \textit{Entropy}, \textbf{19}, 391; \texttt{DOI:10.3390/e19080391} .
\item[ ]

Terenin A, Magnusson M, Jonsson L, and Draper D (2018). P\'olya Urn Latent Dirichlet Allocation: A Doubly Sparse Massively Parallel Sampler. \textit{IEEE Transactions on Pattern Analysis and Machine Intelligence}, \textbf{41}, 1709--1719, \texttt{DOI:10.1109/TPAMI.2018.2832641} .

\end{itemize}

Address to which the proofs should be sent (electronic, not postal, transmission is greatly preferred):

\begin{tabular}{l}

\hspace*{0.2in} Professor David Draper \\
\hspace*{0.2in} Department of Statistics \\
\hspace*{0.2in} Baskin School of Engineering \\
\hspace*{0.2in} University of California \\
\hspace*{0.2in} 1156 High Street \\
\hspace*{0.2in} Santa Cruz CA 95064 USA \vspace*{0.1in} \\

\hspace*{0.2in} phone +1 831 345 5902 \\

\hspace*{0.2in} email draper@ucsc.edu

\end{tabular}

\end{document}